\documentclass[12pt]{amsproc}%
\usepackage{amssymb}
\usepackage{amsmath}
\usepackage{amsfonts}
\usepackage{graphicx}%
\begin{document}
\title{ Zero-norm states and stringy symmetries }
\author{Chuan-Tsung Chan}
\address{Physics Division, National Center for Theoretical Sciences, Hsinchu, Taiwan, R.O.C.}
\email{ctchan@phys.cts.nthu.edu.tw}
\author{Pei-Ming Ho}
\address{Department of Physics, National Taiwan University, Taipei, Taiwan, R.O.C.}
\email{pmho@ntu.edu.tw}
\author{Jen-Chi Lee}
\address{Department of Electrophysics, National Chiao-Tung University, Hsinchu, Taiwan,
R.O.C. (\textbf{Corresponding Author})\\
 }
\email{jcclee@cc.nctu.edu.tw}
\author{Shunsuke Teraguchi}
\address{Department of Physics, National Taiwan University and National Center for\\
Theoretical Sciences, Hsinchu, Taiwan, R.O.C.}
\email{teraguch@phys.ntu.edu.tw}
\author{Yi-Yang}
\address{Department of Electrophysics, National Chiao-Tung University, Hsinchu, Taiwan, R.O.C.}
\email{yiyang@mail.nctu.edu.tw}
\date{\today }

\begin{abstract}
We identify spacetime symmetry charges of 26D open bosonic string theory from
an infinite number of zero-norm states (ZNS) with arbitrary high spin in the
old covariant first quantized string spectrum. We give various evidences to
support this identification. These include massive sigma-model calculation,
Witten string field theory calculation, 2D string theory calculation and, most
importantly, three methods of high-energy stringy scattering amplitude
calculations. The last calculations explicitly prove Gross's conjectures in
1988 on high energy symmetry of string theory.

\end{abstract}
\maketitle

\section{\bigskip Introduction and Overview}

One of the fundamental issues of string theory is the spacetime symmetry of
the theory. It has long been believed that string theory consists of a huge
hidden symmetry or Ward identities. This is strongly suggested by the
ultraviolet finiteness of quantum string theory, which contains no free
parameter and an infinite number of states. In a local quantum field theory, a
symmetry principle was postulated, which can be used to determine the
interaction of the theory. In string theory, on the contrary, it is the
interaction, prescribed by the very tight quantum consistency conditions due
to the extendedness of string, which determines the form of the symmetry.

Historically, the first key progress to understand symmetry of string theory
is to study the high energy, fixed angle behavior of string scattering
amplitude \cite{GM, Gross, GrossManes}. This is strongly motivated by the
spontaneously broken symmetries in gauge field theories which are hidden at
low energy, but become evident in the high-energy behavior of the theory.
There are two main conjectures of Gross's \cite{Gross} pioneer work on this
subject. The first one is the existence of an infinite number of linear
relations among the scattering amplitudes of different string states that are
valid order by order in string perturbation theory at high energies. The
second is that this symmetry is so powerful as to determine the scattering
amplitudes of all the infinite number of string states in terms of the dilaton
(tachyon for the case of open string) scattering amplitudes. However, the
symmetry charges of his proposed stringy symmetries were not understood and
the proportionality constants between scattering amplitudes of different
string states were not calculated.

The second key to uncover the fundamental symmetry of string theory was
zero-norm states (ZNS) in the old covariant first quantized (OCFQ) string
spectrum. It was proposed that \cite{Lee} spacetime symmetry charges of string
theory orginate from an infinite number of ZNS with arbitrary high spin in the
spectrum. In the context of $\sigma$-model approach of string theory, massive
inter-particle symmetries were calculated by using two types of ZNS. Some
implications of the corresponding stringy Ward identities on the scattering
amplitudes were discussed in \cite{JCLee,MassHeter}. It was recently realized
that \cite{KaoLee, CLYang} these symmetries can be reproduced from gauge
transformation of Witten string field theory (WSFT) \cite{Witten} after
imposing the no ghost conditions. It is important to note that these
symmetries exist only for D=26 thanks to type two ZNS, which is zero-norm only
when D=26. On the other hand, ZNS were also shown \cite{ChungLee} to carry the
spacetime $\omega_{\infty}$ symmetry \cite{Winfinity} charges of 2D string
theory \cite{2Dstring}. This is in parallel with the work of Ref \cite{Ring}
where the ground ring structure of ghost number zero operators was identified
in the BRST quantization. Other approaches of stringy symmetries include the
studies of the high-energy, fixed momentum transfer regime \cite{Regge}, the
Hagedorn transition at high temperature \cite{Hagedorn}, vertex operator
algebra for compatified spacetime or on a lattice \cite{Moore}, group
theoretic approach of string \cite{West} and by taking the tensionless limit
of the worldsheet theory \cite{WS}. Despite all these developments, there
seemed to be not much progress on this important subject.

Recently high-energy Ward identities derived from the decoupling of 26D open
bosonic string ZNS, which combines the previous two key ideas of probing
stringy symmetry, were used to explicitly prove Gross's two conjectures
\cite{ChanLee1,ChanLee2, CHL, CHLTY1,CHLTY2}. An infinite number of linear
relations among high energy scattering amplitudes of different string states
were derived. Moreover, these linear relations can be used to fix the
proportionality constants among high energy scattering amplitudes of different
string states algebraically at each fixed mass level. Exactly the same results
can also be obtained by two other calculations, the Virasoro constraint
calculation and the saddle-point calculation. Thus there is only one
independent component of high energy scattering amplitude at each fixed mass
level. Based on this independent component of high energy scattering
amplitude, one can then derive the general formula of high energy scattering
amplitude for four arbitrary string states, and express them in terms of that
of tachyons. This completes the general proof of Gross's two conjectures on
high-energy symmetry of string theory stated above.\ Incidentally, it was
important to discover \cite{ChanLee1,ChanLee2, CHL} that the result of
saddle-point calculation in Refs \cite{GM, Gross, GrossManes} was inconsistent
with high energy stringy Ward identities of ZNS calculation in Refs
\cite{ChanLee1,ChanLee2, CHL}. A corrected saddle-point calculation was given
in Ref \cite{CHL}, where the missing terms of the calculation in Refs
\cite{GM, Gross, GrossManes} were identified to recover the stringy Ward identities.

\section{\bigskip Zero-norm state calculations}

In this section, we review the calculations of string symmetries from ZNS
without taking the high-energy limit. In the OCFQ spectrum of 26D open bosonic
string theory, the solutions of physical states conditions include
positive-norm propagating states and two types of ZNS. The latter are%
\begin{equation}
\text{Type I}:L_{-1}\left\vert x\right\rangle ,\text{ where }L_{1}\left\vert
x\right\rangle =L_{2}\left\vert x\right\rangle =0,\text{ }L_{0}\left\vert
x\right\rangle =0; \tag{1}%
\end{equation}%
\begin{equation}
\text{Type II}:(L_{-2}+\frac{3}{2}L_{-1}^{2})\left\vert \widetilde
{x}\right\rangle ,\text{ where }L_{1}\left\vert \widetilde{x}\right\rangle
=L_{2}\left\vert \widetilde{x}\right\rangle =0,\text{ }(L_{0}+1)\left\vert
\widetilde{x}\right\rangle =0. \tag{2}%
\end{equation}
While type I states have zero-norm at any space-time dimension, type II states
have zero-norm \emph{only} at D=26. Some explicit solutions of ZNS can be
found in \cite{ZNS}. In the $\sigma$-model approach of string theory, a
spacetime symmetry transformation $\delta\Phi$ for a background field $\Phi$
can be generated by \cite{EO}
\begin{equation}
T_{\Phi}+\delta T=T_{\Phi+\delta\Phi}, \tag{3}%
\end{equation}
where $T_{\Phi}$ is the worldsheet energy momentum tensor with background
fields $\Phi$ and $T_{\Phi+\delta\Phi}$ is the new energy momentum tensor with
new background fields $\Phi+\delta\Phi$. It was shown that \cite{Lee} for each
ZNS, one can construct a $\delta T$ such that Eq.(3) is satisfied to some
order of weak field approximation in the $\beta$ function calculation. In
constrast to the usual $\sigma$-model loop expansion (or $\alpha^{\prime}$
expansion), which is nonrenormalizable for the massive background field, it
turns out that weak field approximation is the more convenient expansion to
deal with massive background field. An inter-particle symmetry transformation
for two high spin states at mass level $M^{2}=4$, for example, can be
generated \cite{Lee}%
\begin{equation}
\delta C_{(\mu\nu\lambda)}=(\frac{1}{2}\partial_{(\mu}\partial_{\nu}%
\theta_{\lambda)}-2\eta_{(\mu\nu}\theta_{\lambda)}),\delta C_{[\mu\nu
]}=9\partial_{\lbrack\mu}\theta_{\nu]}, \tag{4}%
\end{equation}
where $\partial_{\nu}\theta^{\nu}=0,(\partial^{2}-4)\theta^{\nu}=0$ are the
on-shell conditions of the mixed type I and type II $D_{2}$ vector ZNS%

\begin{equation}
|D_{2}\rangle=[(\frac{1}{2}k_{\mu}k_{\nu}\theta_{\lambda}+2\eta_{\mu\nu}%
\theta_{\lambda})\alpha_{-1}^{\mu}\alpha_{-1}^{\nu}\alpha_{-1}^{\lambda
}+9k_{\mu}\theta_{\nu}\alpha_{-2}^{[\mu}\alpha_{-1}^{\nu]}-6\theta_{\mu}%
\alpha_{-3}^{\mu}]\left\vert 0,k\right\rangle ,\text{ \ }k\cdot\theta=0,
\tag{5}%
\end{equation}
and $C_{(\mu\nu\lambda)}$ and $C_{[\mu\nu]}$ are the background fields of the
symmetric spin-three and antisymmetric spin-two states respectively at the
mass level $M^{2}=4$. It is important to note that the decoupling of $D_{2}$
vector zero-norm state implies simultaneous change of both $C_{(\mu\nu
\lambda)}$ and $C_{[\mu\nu]}$ , thus they form a gauge multiplet. In WSFT, one
can rederive \cite{KaoLee, CLYang} Eq.(4) from the linearized off-shell gauge
transformation%
\begin{equation}
\delta\Phi=Q_{\text{B}}\Lambda\tag{6}%
\end{equation}
after imposing the no ghost conditions.

Another evidence to support ZNS as the origin of symmetry charge was
demonstrated for the 2D string theory. The spacetime symmetry of 2D string was
known to be the $w_{\infty}$ algebra \cite{Winfinity}%
\begin{equation}
\int\frac{dz}{2\pi i}\psi_{J_{1}M_{1}}^{+}(z)\psi_{J_{2}M_{2}}^{+}%
(0)=(J_{2}M_{1}-J_{1}M_{2})\psi_{(J_{1}+J_{2}-1)(M_{1}+M_{2})}^{+}(0) \tag{7}%
\end{equation}
generated by the discrete Polyakov states $\psi_{JM}^{+}$. An equivalent
algebraic structure was the ground ring \cite{Ring}
\begin{equation}
Q_{J_{1},M_{1}}\cdot Q_{J_{2},M_{2}}=Q_{J_{1}+J_{2},M_{1}+M_{2}} \tag{8}%
\end{equation}
proposed by Witten. Alternatively, one can explicitly construct a set of
discrete ZNS $G_{JM}^{+}$ and show that they form a $w_{\infty}$ algebra
\cite{ChungLee}%
\begin{equation}
\int\frac{dz}{2\pi i}G_{J_{1}M_{1}}^{+}(z)G_{J_{2}M_{2}}^{+}(0)=(J_{2}%
M_{1}-J_{1}M_{2})G_{(J_{1}+J_{2}-1)(M_{1}+M_{2})}^{+}(0). \tag{9}%
\end{equation}
This seems to strongly suggest that ZNS are closely related to the spacetime
symmetry of string theory.

\section{High energy ZNS calculations}

Recently a further evidence to support ZNS as the spacetime symmetry charge of
string theory was obtained by taking the high-energy, fixed angle limit of
stringy Ward identities derived from the decoupling of ZNS on the scattering
amplitudes. The two conjectures of Gross stated above were then explicitly
proved. At mass level $M^{2}=4$, for example, the stringy Ward identities for
four-point functions derived from the decoupling of four zero-norm states were
calculated to be \cite{JCLee}
\begin{equation}
k_{\mu}\theta_{\nu\lambda}\mathcal{T}_{\chi}^{(\mu\nu\lambda)}+2\theta_{\mu
\nu}\mathcal{T}_{\chi}^{(\mu\nu)}=0, \tag{10}%
\end{equation}%
\begin{equation}
(\frac{5}{2}k_{\mu}k_{\nu}\theta_{\lambda}^{\prime}+\eta_{\mu\nu}%
\theta_{\lambda}^{\prime})\mathcal{T}_{\chi}^{(\mu\nu\lambda)}+9k_{\mu}%
\theta_{\nu}^{\prime}\mathcal{T}_{\chi}^{(\mu\nu)}+6\theta_{\mu}^{\prime
}\mathcal{T}_{\chi}^{\mu}=0, \tag{11}%
\end{equation}%
\begin{equation}
(\frac{1}{2}k_{\mu}k_{\nu}\theta_{\lambda}+2\eta_{\mu\nu}\theta_{\lambda
})\mathcal{T}_{\chi}^{(\mu\nu\lambda)}+9k_{\mu}\theta_{\nu}\mathcal{T}_{\chi
}^{[\mu\nu]}-6\theta_{\mu}\mathcal{T}_{\chi}^{\mu}=0, \tag{12}%
\end{equation}%
\begin{equation}
(\frac{17}{4}k_{\mu}k_{\nu}k_{\lambda}+\frac{9}{2}\eta_{\mu\nu}k_{\lambda
})\mathcal{T}_{\chi}^{(\mu\nu\lambda)}+(21k_{\mu}k_{\nu}+9\eta_{\mu\nu
})\mathcal{T}_{\chi}^{(\mu\nu)}+25k_{\mu}\mathcal{T}_{\chi}^{\mu}=0, \tag{13}%
\end{equation}
where $\theta_{\mu\nu}$ is symmetric, transverse and traceless, and
$\theta_{\lambda}^{\prime}$ and $\theta_{\lambda}$ are transverse vectors.
These are polarizations of zero-norm states. $\mathcal{T}_{\chi}^{\prime}s$ in
Eqs.(10)-(13) are $\chi$-th order string-loop amplitudes. In each equation, we
have chosen, say, the second vertex $V_{2}(k_{2})$ in the correlation function
to be constructed from zero-norm states at the mass level $M^{2}$ $=4$ and
$k_{\mu}\equiv k_{2\mu}$. The tensor index of the other three string vertex
were suppressed in Eq.(10)-(13). By using Eqs.(10)-(13), the leading order
Ward identities of Eqs.(10)-(13) in the $\frac{1}{E^{2}}$ expansions were
calculated to be (we drop loop order $\chi$ here to simplify the notation)
\cite{ChanLee1,ChanLee2}
\begin{equation}
\mathcal{T}_{LLT}+\mathcal{T}_{(LT)}=0, \tag{14}%
\end{equation}%
\begin{equation}
10\mathcal{T}_{LLT}+\mathcal{T}_{TTT}+18\mathcal{T}_{(LT)}=0, \tag{15}%
\end{equation}%
\begin{equation}
\mathcal{T}_{LLT}+\mathcal{T}_{TTT}+9\mathcal{T}_{[LT]}=0. \tag{16}%
\end{equation}
In the above equations, we have defined the normalized polarization vectors of
the second string vertex to be $e_{P}=\frac{1}{m_{2}}(E_{2},\mathrm{k}%
_{2},0)=\frac{k_{2}}{m_{2}},$ $e_{L}=\frac{1}{m_{2}}(\mathrm{k}_{2},E_{2},0)$
and $e_{T}$ $=(0,0,1)$ in the CM frame contained in the plane of scattering.
It turns out that $e_{P}$ approaches $e_{L}$ in the high-energy limit. A
simple calculation shows that, in contrast to Eq.(4) which is valid to all
energies, the linear relations among the high-energy scattering amplitudes are
\cite{ChanLee1,ChanLee2}
\begin{equation}
\mathcal{T}_{TTT}:\mathcal{T}_{LLT}:\mathcal{T}_{(LT)}:\mathcal{T}%
_{[LT]}=8:1:-1:-1. \tag{17}%
\end{equation}
Note that the underlying high-energy limit of zero-norm states in the
equations above are no longer of zero norm. That is why we will be able to
obtain nontrivial relations among physically inequivalent particles. For the
case of string-tree level $\chi=1$ with one tensor $V_{2}$ and three tachyons
$V_{1,3,4}$, all four scattering amplitudes in Eq.(17) were calculated to be
$\mathcal{T}_{TTT}=-8E^{9}\sin^{3}\phi_{CM}\mathcal{T}(3)=8\mathcal{T}%
_{LLT}=-8\mathcal{T}_{(LT)}=-8\mathcal{T}_{[LT]}$ , where
\begin{align}
\mathcal{T}(n)  &  =\sqrt{\pi}(-1)^{n-1}2^{-n}E^{-1-2n}(\sin\frac{\phi_{CM}%
}{2})^{-3}(\cos\frac{\phi_{CM}}{2})^{5-2n}\nonumber\\
&  \times\exp(-\frac{s\ln s+t\ln t-(s+t)\ln(s+t)}{2}). \tag{18}%
\end{align}
In Eq.(18), $\phi_{CM}$ is the center of momentum scattering angle, $s,t$ and
$u$ are the Mandelstam variables and $M^{2}=2(n-1)$.

For the scattering amplitudes of general mass levels, one first notes that the
only states that will survive in the high energy limit at level $M^{2}=2(n-1)$
are of the form%
\begin{equation}
\left\vert n,2m,q\right\rangle \equiv(\alpha_{-1}^{T})^{n-2m-2q}(\alpha
_{-1}^{L})^{2m}(\alpha_{-2}^{L})^{q}\left\vert 0,k\right\rangle . \tag{19}%
\end{equation}
The next step is to use the decoupling of two types of high-energy ZNS
\begin{equation}
L_{-1}\left\vert n-1,2m-1,q\right\rangle \simeq M\left\vert
n,2m,q\right\rangle +(2m-1)\left\vert n,2m-2,q+1\right\rangle , \tag{20}%
\end{equation}%
\begin{equation}
L_{-2}\left\vert n-2,0,q\right\rangle \simeq\frac{1}{2}\left\vert
n,0,q\right\rangle +M\left\vert n,0,q+1\right\rangle , \tag{21}%
\end{equation}
to deduce that \cite{CHLTY1,CHLTY2}
\begin{equation}
\mathcal{T}^{(n,2m,q)}=\left(  -\frac{1}{M}\right)  ^{2m+q}\left(  \frac{1}%
{2}\right)  ^{m+q}(2m-1)!!\mathcal{T}^{(n,0,0)}. \tag{22}%
\end{equation}
For the case of general four tensor scattering amplitude, one has, in the high
energy limit,%
\begin{equation}
<V_{1}V_{2}V_{3}V_{4}>=%
{\textstyle\prod_{i=1}^{4}}
(-\frac{1}{M_{i}})^{2m_{i}+q_{i}}(\frac{1}{2})^{m_{i}+q_{i}}(2m_{i}%
-1)!!\mathcal{T}_{n_{1}n_{2}n_{3}n_{4}}^{T^{1}\cdot\cdot T^{2}\cdot\cdot
T^{3}\cdot\cdot T^{4}\cdot\cdot}, \tag{23}%
\end{equation}
which is calculated algebraically by the decoupling of high-energy ZNS and is
thus valid to all string-loop order. $\mathcal{T}_{n_{1}n_{2}n_{3}n_{4}%
}^{T^{1}\cdot\cdot T^{2}\cdot\cdot T^{3}\cdot\cdot T^{4}\cdot\cdot}$ in
Eq.(23), the generalization of $\mathcal{T}_{TTT}$ at mass level $M^{2}=4$, is
the only independent high-energy scattering amplitudes at level $(n_{1}%
,n_{2},n_{3},n_{4})$ and was calculated at tree level to be \cite{ChanLee1,
CHL}%
\begin{equation}
\mathcal{T}_{n_{1}n_{2}n_{3}n_{4}}^{T^{1}\cdot\cdot T^{2}\cdot\cdot T^{3}%
\cdot\cdot T^{4}\cdot\cdot}=[-2E^{3}\sin\phi_{CM}]^{\Sigma n_{i}}%
\mathcal{T}(\Sigma n_{i})=2\sqrt{\pi}e^{n-4}(stu)^{\frac{n-3}{2}}e^{-\frac
{1}{2}(s\ln s+t\ln t+u\ln u)}, \tag{24}%
\end{equation}
where $n_{i}$ is the number of $T^{i}$ of the $i-th$ vertex operators,
$n=\sum_{i=1}^{4}n_{i}$ and $T^{i}$ is the transverse direction of the $i-th$ particle.

The decoupling of two types of high-energy ZNS in Eqs.(20) and (21) can be
shown to be equivalent to the decoupling of two types of zero-norm states in
Eqs.(1) and (2) in the high-energy limit. It is interesting to see that the
second term $\frac{3}{2}L_{-1}^{2}\left\vert \widetilde{x}\right\rangle $ of
type II zero-norm states in Eq.(2) can be dropped out in the high-energy limit
without affecting the final result. This hints at a \textquotedblleft
dual\textquotedblright\ calculation, the high-energy limit of the Virasoro
constraints $L_{1}\left\vert \psi\right\rangle =L_{2}\left\vert \psi
\right\rangle =0$, to derive Eq.(22) \cite{CHLTY1,CHLTY2}. Finally, a
saddle-point calculation \cite{CHL, CHLTY1,CHLTY2} can be developed for the
string-tree level $\chi=1$ scattering amplitudes with one tensor $V_{2}$ and
three tachyons $V_{1,3,4}$ to explicitly justify Eq.(22).

For the case of 2D string theory, The discrete 2D ZNS $G_{JM}^{+}$ constructed
in \cite{ChungLee} can be shown \cite{CHLTY1} to approach to the discrete
Polyakov states $\psi_{JM}^{+}$ in the high-energy limit. Thus the $w_{\infty
}$ algebra constructed in Eq.(9) can be identified to that of Eq.(7) in the
high-energy limit.

\end{document}